# The effect of the COVID-19 pandemic on gendered research productivity and its correlates

Eunrang Kwon[1], Jinhyuk Yun[2]*, Jeong-han Kang[1]*

[1]Yonsei University, Department of Sociology, Seoul, Korea

[2]Soongsil University, School of AI Convergence, Seoul, Korea

* Corresponding authors: Correspondence to Jinhyuk Yun(jinhyuk.yun@ssu.ac.kr) or Jeong-han Kang (jhk55@yonsei.ac.kr).

## Abstract

Female researchers may have experienced more difficulties than their male counterparts since the COVID-19 outbreak because of gendered housework and childcare. Using Microsoft Academic Graph data from 2016 to 2020, this study examined how the proportion of female authors in academic journals on a global scale changed in 2020 (net of recent yearly trends). We observed a decrease in research productivity for female researchers in 2020, mostly as first authors, followed by last author position. Female researchers were not necessarily excluded from but were marginalised in research. We also identified various factors that amplified the gender gap by dividing the authors' backgrounds into individual, organisational and national characteristics. Female researchers were more vulnerable when they were in their mid-career, affiliated to the least influential organisations, and more importantly from less gender-equal countries with higher mortality and restricted mobility as a result of COVID-19.

## Introduction

The COVID-19 pandemic has changed the way we live everywhere, and academia has not been immune to these changes. For instance, research productivity declined during the pandemic, particularly for female researchers [1–5]. It is evident that everyone has been affected by the pandemic, yet the degree of disturbance may differ. To illustrate, as of 11th October 2021, four of the five countries with the highest COVID-19 death tolls [6] were developing or emerging economies identified by the International Monetary Fund (IMF) [7]. Moreover, this influence diverges according to scholars' environment and characteristics (such as their field of study (FOS), social class, and country of domicile); thus, the decline in scientific productivity may not occur equally for researchers. To understand this disparity during the pandemic, scholars have explored the impact of social background on research activity [3,5,8,9]. They found that academic minors, particularly women, suffer more. However, despite the significant implications of such studies, they have generally focused on gauging impact within a small sub-community rather throughout the academic community overall. This necessitates complementary research with a more quantitative and unbiased focus. Here, on the basis of large-scale bibliographic data, we attempt a systematic and quantitative comprehension of the unequal impact of the pandemic on scientific activity regarding the status of researchers.

We hypothesise that researcher vulnerability should be reflected in research productivity and suggest an extensive collection of scientific publications at a pan-disciplinary scale might be suitable for gauging their impact. Microsoft Academic Graph (MAG) provides information with metadata regarding millions of scientific publications, including journal papers, conference proceedings, and online preprints. Here, we focus on the impact of gender on research activity as a personal factor that is vulnerable to the effects of the pandemic. In particular, the research productivity of female researchers may have decreased more than that of the male researchers because of the unequal division of housework and childcare between husband and wife, which has intensified since the outbreak of COVID-19 [10,11]. Because bibliographical datasets do not identify the gender of authors, we have identified them with a statistics-based gender classifier (see Methods).

Female academics have established their position in all subject areas but have not yet achieved full gender equity. In most cultures, unpaid housework and more subtle emotion work[12] at home is disproportionately imposed on women due to conventional gender roles; thus, these additional burdens may result in unequal sacrifice [13,14]. To test the hypothesis, we examined the gendered impact of research activity according to social and academic background at three levels: individual (reflecting career stage and academic prestige); organisational (reflecting institutional prestige); and country (reflecting living environment, such as gender equality and the severity of COVID-19 at the national level).

One clear advantage of investigating such pan-academic scale data is that the effects of the pandemic are felt across the entire academic community and are not limited to a specific subgroup (on which previous studies have focused). The study used all publications in the MAG from 2016 to 2020, regardless of either the journal or discipline. We combined this large-scale bibliographic metadata with scholars' social and academic backgrounds. The academic age and author and affiliation *h*-indices were selected as the representative characteristics of individuals and their academic organisations. Meanwhile, gender equality and total cases of infection/mortality per million (COVID-19) were chosen as the national statistics. We grouped papers into four levels for each variable and explored gender impact by investigating the change in difference (CID) of female authors' proportions between adjacent years (see Methods). CID can be conceived as a variant of difference-in-difference (DID) between two groups [15,16] that cancels any common effect on two differences. This can capture net change in 2020 beyond the constant increase/decrease of female authors' participation during recent years. Our analysis demonstrates that the pandemic had a greater negative impact on females than males, as presented in the negative CID in 2020 compared with previous years. The analysis (taking into account authors' surrounding factors) further reveals under which conditions women in academia were more seriously affected in terms of their career stage, affiliated institution, and country of work.

## Relevant Studies

In this section, we provide a concise review of the studies and debates related to the gendered effect of the COVID-19 pandemic in academia. The outbreak of COVID-19 is an unprecedented disaster on a global scale and has led to the explosion of "COVID science" [17]. However, there have been limited studies on changes in academia globally as a result of the pandemic. Meanwhile, considerable concern has arisen in studies on the decline of research productivity of female researchers in various academic fields. Many countries continue to practice social distancing and some have even enforced lockdowns to prevent the spread of the virus, resulting in restricted mobility [18]. Research facilities have been closed, and remote work has been widely implemented to avoid face-to-face contact. At the same time, children have been kept home without going to school or childcare. This has resulted in an additional burden of housework and childcare that is not equally distributed among males and females in most households. Females have had to undertake more unpaid family work [19], which likely yields a decrease in research productivity compared with male counterparts. This may reasonably be assumed to be because of conventional gender roles in many cultures [14,20].

While insufficient time has been spent on research throughout the academic community, it has been reported that female researchers have experienced career disruption due to an increased burden of childcare and housework compared to male counterparts. This suggests a possible long-term negative effect on female researchers as a result of COVID-19 [10,11,21]. This finding suggests that an increase in childcare responsibilities at home is a cause of the decrease in research productivity in the STEM fields as well as others such as linguistics and sociology [3,22]. Another study demonstrated that lockdown has resulted in a loss of female economic productivity—also reflected in lower participation rates for females among new research related to COVID-19 [23].

In summary, the impact of COVID-19 on research productivity may vary depending on the stage of life course for female researchers, especially for those who are parents [22,24]. Such gendered effects are also observed in online environments. To illustrate, women are less successful than men in disseminating their research and its impact [25], while gendered tie formation in co-authorship networks is associated with online success for males but not for females. Preprints in the online repository Social Science Research Network (SSRN) show that female research productivity dropped more than that of males, despite the total research productivity increasing after 10 weeks of lockdown in the United States [10]. In short, even though some studies focus on the gendered impact of the pandemic, the main evidence is based on a rather small subset of academia and commonly neglects to estimate the impact of authors' environment that may either promote or disrupt scientific productivity.

Some studies report the influence of the author's role in research during the pandemic, although these only test for rather a small subset (such as a single journal or FOS). One significant observation was that fewer females were presented as first author and had fewer submission rates compared with males [1,2,4]. Other studies point out that the underestimated contribution of female authors leads to unfair distribution of authorship [26–28]. Meanwhile, women as last author were found to be inconsistent across studies [1,2,4,8].

Taken together, these findings indicate that it is indispensable to differentiate author roles if we aim to sharply analyse gendered research productivity. For instance, in the physical sciences and biology, the first author is usually an early-career researcher who primarily implements the research, whereas the last author mainly takes a mentoring role that steers the research [29]. Therefore, for a single research paper, the first author can be considered a person who devotes the most effort among the authors; from this we can hypothesise that leading authors have been influenced more negatively than the other authors by heavier household duties and mental distractions during the COVID-19 pandemic [24].

In addition to author role, we suggest some additional factors that may interact with gender in affecting research productivity: academic career stage, prestige, and national contexts concerning gender and epidemiological situations. First, for the interacting effect of career stage, we hypothesise that female researchers' productivity will be more affected by the pandemic when their career stage is more likely to be confounded with parenthood. For instance, early-career female authors have experienced a decrease in research productivity during the pandemic, although the influence on female authorship overall was inconclusive in the field of medicine [8]. In the United States, highly educated women usually begin motherhood in their 30s [30]. Because the earliest age for the first publication can be determined to be around the mid-20s, we can assume that mid-career researchers are likely to have minor children who need parental care [31].

Another possible protective effect of academic prestige is how the author's profile in academia affects vulnerability. Two competing hypotheses are possible. First, less prestigious female authors may be more vulnerable to the pandemic because of insufficient resources and support. Second, more prestigious authors may be more vulnerable due to harsher competition with comparably prestigious male academic counterparts.

Females are known to be underrepresented in the academic world, especially in top positions. For example, females tend to be evaluated by performance less favourably than males, which influences the small number of leading positions available for females [32]. Similarly, in America, female academics do not obtain tenure and promotion at the same rate as men in similar fields with similar academic credentials [33]. Moreover, females still report stress from lack of administrative support, current economic conditions, bias of tenured colleagues, and family responsibilities [34]. The glass ceiling and the maternal wall, formed by gender stereotypes, demote their position and shorten their tenure [34]. In addition, females are deprived of research time because they tend to be assigned more 'feminine' tasks, such as student counselling and local arrangement of conferences.

Although there is no absolute standard for scientific prestige of either academics or their affiliations, the citation-based bibliometric indicators, such as number of citations, papers with the highest number of citations, and scientific proximity, are usually employed [35]. These measures are generally focused on either of two points: i) productivity of authors and ii) citation impact of their publications. The h-index (an increasing function of time) is a well-known measure that can embrace both points to quantify the research output of individual scientists [36], and can also be applied as an affiliation-level metric [37], although there is a bias towards mature scientists [38]. Therefore, it can be an indicator of both career longevity and academic impact. In other words, a high h-index indicates a stable research environment that secures authors from external threats.

Gendered culture in laboratory-based and academic occupations can be another intervening factor. The survey demonstrates that physical research time in disciplines requiring laboratory work (such as biology, chemistry, and engineering) decreased sharply after COVID-19, while there was little impact for dry laboratories [5]. One exception is time spent on medical research related to COVID-19 [5]. With a higher care burden for female researchers (as already discussed) [10,11], the research productivity of female researchers in STEM decreased after the outbreak of the pandemic [3,8]. Females who are expected to conform to traditional gender roles in male-dominated organisations consider themselves excluded and marginalised from the majors [39]. Therefore, female researchers in STEM fields are in more critical environments than males during the pandemic. While it may be extremely difficult to detect gender-related cultures surrounding researchers, this could be approximated by a national gender-equality index for professional occupations [40].

The last interacting factor is the epidemiological risk at the national level. Lockdown and social distancing policies are mostly executed at the national level. Female researchers in a country of higher risk may have been more likely to work from home, absent from the laboratory, and taking care of home-schooling children. Those epidemiological risks can be measured by the cumulative number of infections and deaths by COVID-19 [41].

Here, based on large-scale empirical publication history before and during the first year of the pandemic, we seek to understand the gendered effect of researcher productivity in a quantitative manner.

## Results
### *Changes in Productivity and Collaboration*
Before examining gendered productivity, we present how overall productivity and the degree of collaboration have changed since the COVID-19 outbreak (see Fig. 1). We measure the degree of collaboration by the average number of authors per paper and productivity for each field by the total number of papers in the field. Most fields demonstrate an increased number of papers from 2019 to 2020 (net of common trends since 2018), except for three fields—art, history and philosophy—whose CIDs for number of papers are not positive in 2020 (see Supplementary Fig. 1a for detailed field labels). Also note that no field appears in the third quadrant where both the number of co-authors and the number of papers decreased at the same time in 2020. Therefore, our analysis of the entire year (2020) suggests that collaboration and research productivity in most academic fields increased, arguably as a result of the COVID-19 pandemic (see Supplementary Fig. 1b for further comparisons with previous years).

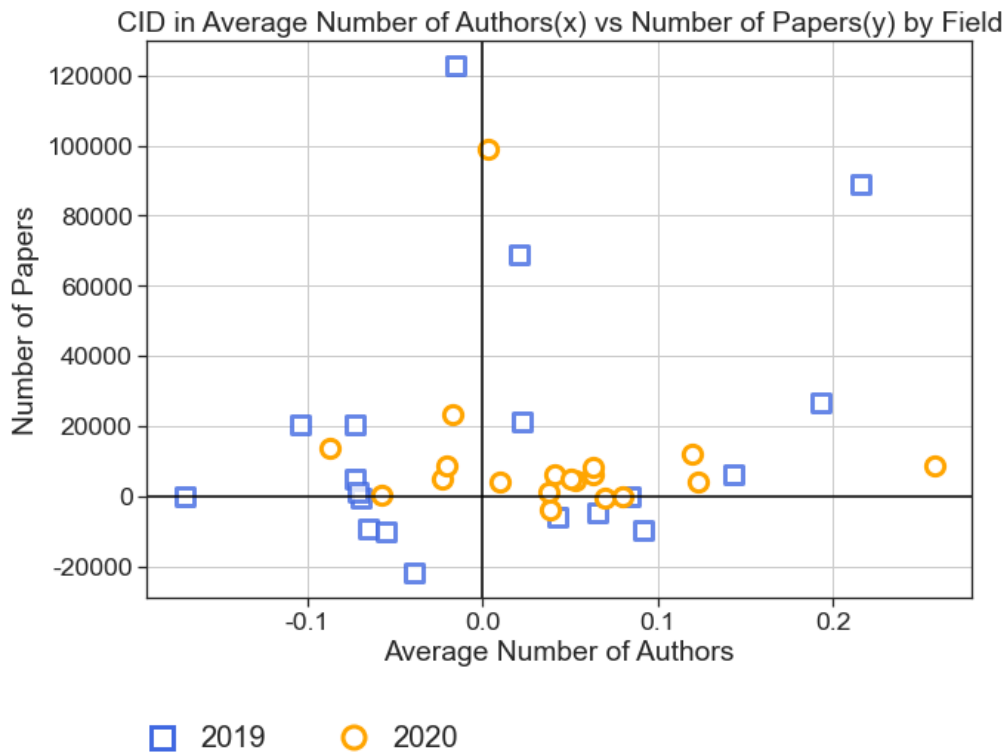

**Figure 1. CID in Average Number of Authors per Paper and Number of Papers by Field.** For all 19 fields of study, we present CIDs (see Methods) in the average number of authors per paper (x-axis) and CIDs in the number of papers (y-axis) for 2020 (orange circles) and 2019 (blue squares). There are no clear patterns for different distributions between two years. CIDs for the number of papers even increased and were mostly distributed above the x-axis. Neither productivity (number of papers) nor collaboration (number of authors) has decreased since the COVID-19 outbreak. Productivity in most fields improved during the pandemic.

Given the increased productivity and collaborations during the first year of the pandemic, we were interested to determined how productive females academic were, depending on their collaborative role. The contribution of female authors to research is often devalued, which leads to cumulative disadvantages in authorship [26]. At the same time, conventional gender stereotypes characterise housework and childcare as female duties. We may assume that female researchers are more challenged than males during the pandemic. [5,9] In many fields, the first author is considered the primary member who conducts the research. They are usually early-career scientists who ask for mentoring from the senior researchers (the last author) [29]. In other words, the first author requires more time on the research than other authors; meanwhile, the last author may invest substantial time in finalising the study, but less than the first author. COVID-19 has led to an abrupt change in society. Under the economically and socially uncertain environment resulting from the pandemic, the first authors in charge of practical analysis may be more vulnerable than other authors because they have to invest the most time among all the authors. Combining the gendered effect and the role of authors, we focus on how much more female first authors have been disadvantaged than those in other authorial positions during the pandemic.

*Female Productivity by Authorship*

To estimate gendered productivity, we first need to identify authors' gender. Unfortunately, the metadata does not include information about the gender of authors; we thus infer the gender using a web-based application that incorporates various census data (see Methods). Also note that the overall participation of female authors has increased in recent years (see Supplementary Fig. 3) [42]. To compensate for this increasing trend, we employed the CID measure, which enables comparison of the yearly variations net of the common effect across recent years (see Methods), with the results presented in Fig. 2.

First, we observe that the change of research productivity is insignificant for female authors post-pandemic when we ignore the role of authors (Fig. 2a). Compared with the recent pre-pandemic period of 2019, the CID (%) has decreased from 0.34 to −0.02. This switch to a decreasing trend is also valid for the non-leading authors who are neither the first nor last authors. When we account for roles, we observe a sharply decreasing tendency of females as the first authors during the pandemic (Fig. 2a). The difference is clear when we compare it with the pre-pandemic years at an increase of CID (%) by 0.28. This pattern is also valid for the last authors, although less salient. The CID of the last authors was positive (=0.07) pre-pandemic, which later became negative in the post-pandemic era (=−0.09).

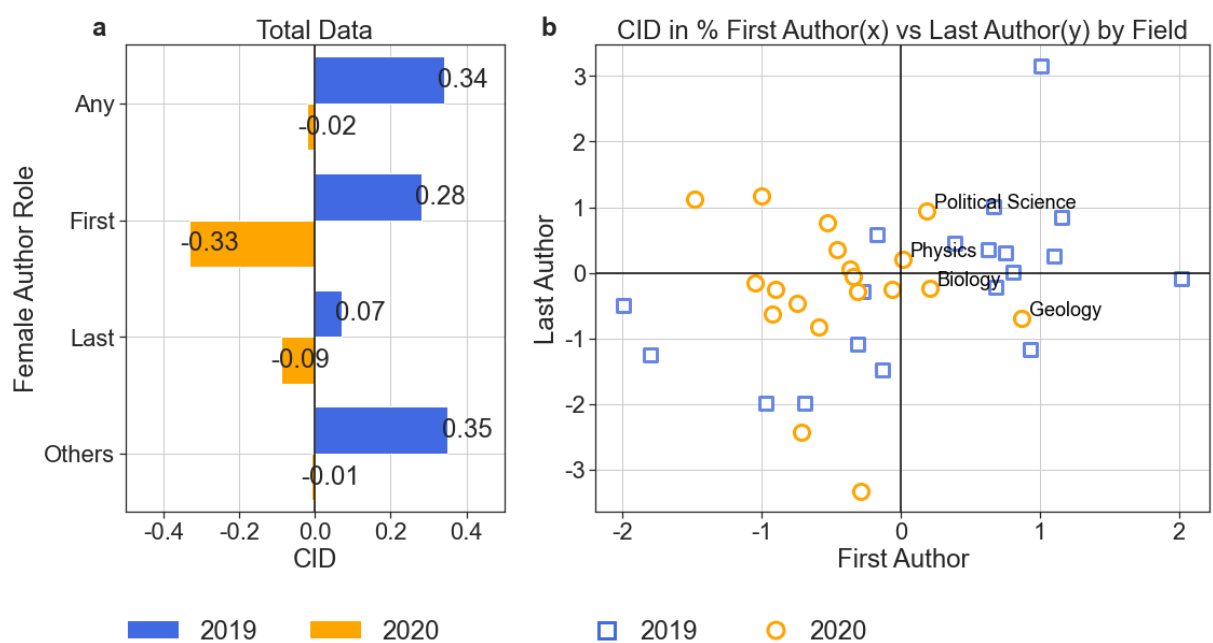

**Figure 2. CID (% female) by author's role and field of study.** The CID for a year estimates how much larger or smaller the change from the previous year is compared with past change to the previous year. **a:** We measured the CID for the percentage of female authors broken down by authors' roles in a paper: the first, the last, and the rest (others). This shows that the % change of female first authors drops most significantly in 2020 by 0.33% (orange bars), reversing the increase in trend up to 2019 (blue bars). Although less dramatic, we also observe decreases in the % change of female last authors, which is unobserved for other author roles. Despite the significant decrease of female authors in first and last authorship, CID for any authorship decreased by only 0.02%, which suggests that the majority of female authors play other roles than either first or last author. **b:** We drew the scatter plot of CIDs between the first and last authors by 19 fields of study post-pandemic at 2020 (orange circle) and pre-pandemic at 2019 (blue square). Compared to evenly scattered CIDs for 2019, CIDs for 2020 tend to be positioned to the left side of the y-axis, except for four fields. In other words, the % of female first authors switched to a decreasing trend in 2020 for 16 out of 19 academic fields. Although less apparent, more than half of CIDs for 2020 are located under the x-axis, which implies a decreased % of last female authors in most fields. In short, what we observe in panel **a** is not attributable to a few fields but common across various fields (for comparing to CIDs between 2018 and 2019, see Supplementary Fig.2).

It could be inferred that the trend may vary for the FOS. Indeed, we find variations of CIDs across the 19 Level 0 FOSs defined by MAG. For instance, the CIDs of political science, biology, and geology are positive for the first authors. However, the CIDs are negative for the majority of FOSs (specifically, 15 of 19), which is consistent with what we observe in Fig. 2a at the aggregate level. The negative CIDs are less apparent for the last authors but hold for over half of the FOSs (12 of 19). Note that about half of FOSs show decreased female participation for both the first and last authors with negative CIDs. In summary, the number of lead female authors decreases in many fields regarding both first and last authors, which is particularly serious for the first authors who execute experiments and write papers themselves. In other words, female authors have tended to take less important roles during the pandemic.

Therefore, although the overall research productivity of female scholars seems not to have significantly decreased, we observe changes in their role during 2020. Female authors are less likely to take leading positions, such as the first and last authors. Considering the reports of inequality in terms of housework and childcare responsibilities [19], we may infer that the pandemic has limited the time and energy that female researchers can devote to research; thus, they may opt for less active roles.

### *Female Productivity by Individual, Organisational, and National Characteristics*

The transition of female authors from active roles to non-active raises an intriguing question as to how much the impact depends on female authors' life stage and academic status. For instance, the presence and age of children can be a critical factor for parenting researchers' productivity. In the United States, highly educated females tend not to start their motherhood until their 30s [43]. Thus, mid-aged researchers are more likely to have minor offspring who need care, resulting in more time being spent on childcare by these women. Unfortunately, the data does not document the age of authors to prove this hypothesis; we thus alternatively employed academic age, which refers to elapsed years from the first paper. We divided papers into four groups by every 10 years of the first author's academic age (see Methods). Calculating the percentage of female first authors and their CIDs, we find clearly decreasing percentage changes for female first authors of all academic age ranges, the highest being for mid-ages in academia (groups 2 and 3 with CID of −0.39 and −0.55, respectively; see Fig. 3a). On average, researchers publish their first paper in their mid-20s [31]; thus, academic age can be estimated as 25 years behind the age of researchers. Groups 2 and 3 correspond to the researchers from their mid-30s to the mid-50s. Because motherhood of highly educated females usually begins after their 30s, it can be inferred that these groups are more likely to have children to care for than other groups. In addition, group 3 (most significantly affected), may have teenage offspring. In summary, this finding supports our hypothesis that childcare affects the productivity of female researchers.

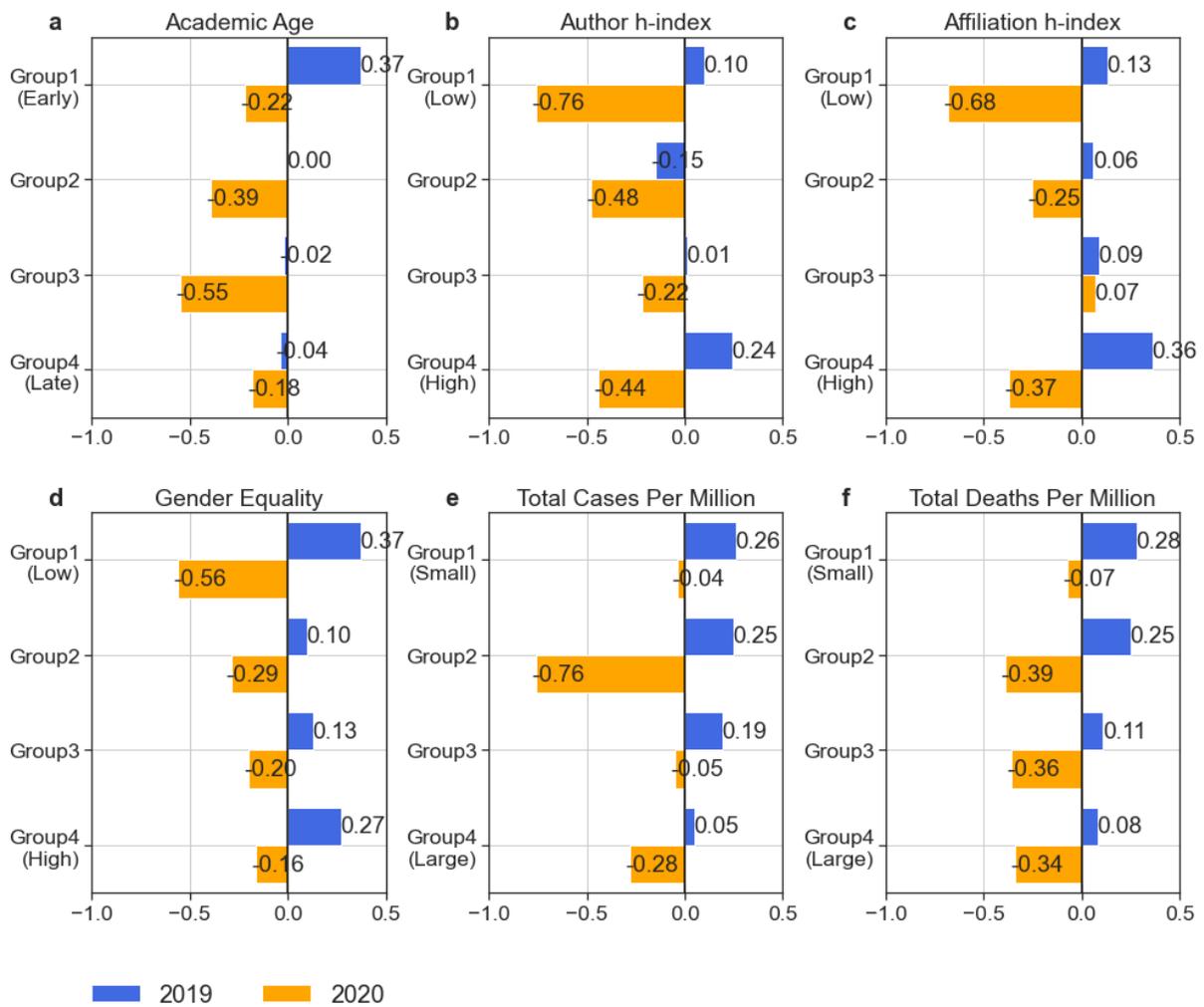

**Figure 3. CID for % female first authors by first author's characteristics.** We grouped the papers into four levels for each of the first author's six characteristics and compared CIDs for % female first authors across the four groups. We considered three individual- and institutional- level characteristics: **a,** academic age since the first publications, **b,** author h-index reflecting authors prestige, **c,** affiliation h-index displaying the status of the affiliations (see Methods), and three country-level statistics, **d,** the level of gender equality (see Methods), and **e**, the total infected cases, and **f,** mortality per million for each country (see Supplementary Fig. 4 for the averaged CID between 2018 and 2019 to compensate for the yearly fluctuation).

Besides academic age, the status of authors and their affiliations can be critical elements of productivity during the pandemic [44]. We used the h-index, which is widely accepted for the author- and affiliation-level performance metrics to gauge prestige. The h-index was high for the prominent authors (and affiliations) with well-established academic positions (see Methods). We found that publications with the least established first authors in group 1 had the biggest fall in percentage change among female first authors, implying that least established females were more severely affected than females in other groups, regarding both authors and their affiliations (Fig 3b and c). This effect decreases as they become more established, except for the most prestigious authors and affiliations in group 4. In sum, the negative CID for the proportion of female first authors shows an inverted U shape across prestige (see Supplementary Fig. 5), both for individual and organisational levels, as a combination of low-prestige vulnerability and high-prestige competition. Previous studies suggest that female faculty experience stresses (such as lack of support and poor economic conditions) and are underrepresented in top positions in academia [32]. The glass ceiling is a barrier to promotion and tenure for female researchers and the assignment of tasks due to gender stereotypes diminishes the available time for female academics to engage in research time [34].

For a more detailed understanding of the decline of female authors' first authorship, we further examined the relationship between the living conditions of their country and the productivity of female authors. Because we hypothesise that an increased burden of housework and childcare means that female authors spend less time on research, overall gender inequality in the country may influence the degree of disadvantage experienced by female researchers (Methods). Indeed, we observed a negative correlation between gender equality and the decline in first authorship for females (Fig. 3d). The decline of CID in 2020 is greatest for group 1 in the most unequal countries (=−0.56), whereas it is lowest for the most equal countries belonging to group 4 (=−0.16). However, we also observed that productivity decreased regardless of the gender equality index during the pandemic.

It may be expected that the severity of COVID-19 in the country of domicile has an influence on female productivity because epidemiological policies tend to be applied at the national level and female researchers will be under the same influence of those policies However, we barely observed consistency with the total infected cases per million (Fig. 3e). The CID in 2020 is the lowest for group 2, or countries with moderate infection rates. The mortality rate shows more consistent results (Fig. 3f). The CIDs in 2020 are high for groups 2–4, the countries with moderate to high mortality rates. In other words, the countries that handle mortality rates well also have the least decrease in female first authorship (see group 1 in Fig. 3f). It seems that mortality rates by COVID-19, rather than infection rates, is a better indicator of living conditions that may influence female disadvantage during the pandemic.

In summary, we found that female first authors' research productivity during the pandemic decreases in the following cases: i) if they are mid-career and likely to have minor offspring, ii) if they are less well academically established, iii) if they live in gender-unequal countries, and iv) unless they live in countries with a very low mortality rate for COVID-19. Note that the infection rate itself does not show a significant correlation with the change in female first authors. Successfully suppressed infections could be both a favourable condition for female productivity and an outcome of intensive social distancing, which increases female household burdens.

### *The Impact of Mobility on Research Productivity*

Many countries have tried to prevent the spread of the COVID-19 by limiting in-person contact. Physical and social distancing, including severe lockdown, has been practiced by many countries. These practices essentially limit personal mobility; thus, mobility can also be considered a COVID-19–related variable, similar to total cases/deaths per million. Both lockdown and distancing certainly reduce the impact of COVID-19 [45], although changes in mobility may vary from country to country because each country has applied different distancing policies. Indeed, there is a correlation between total cases/deaths per million and change in mobility, although this is not completely deterministic (Fig. 4a and b). A social distancing policy typically reduces mobility to workplaces and increases mobility in residential areas, both of which typically increase the burden of females at home because of the closure of school and childcare during the pandemic [19]. Their home-based work overlaps with their children's home-schooling and they cannot consequently dedicate themselves to their research. Therefore, a change in mobility may lead to a decrease in research productivity.

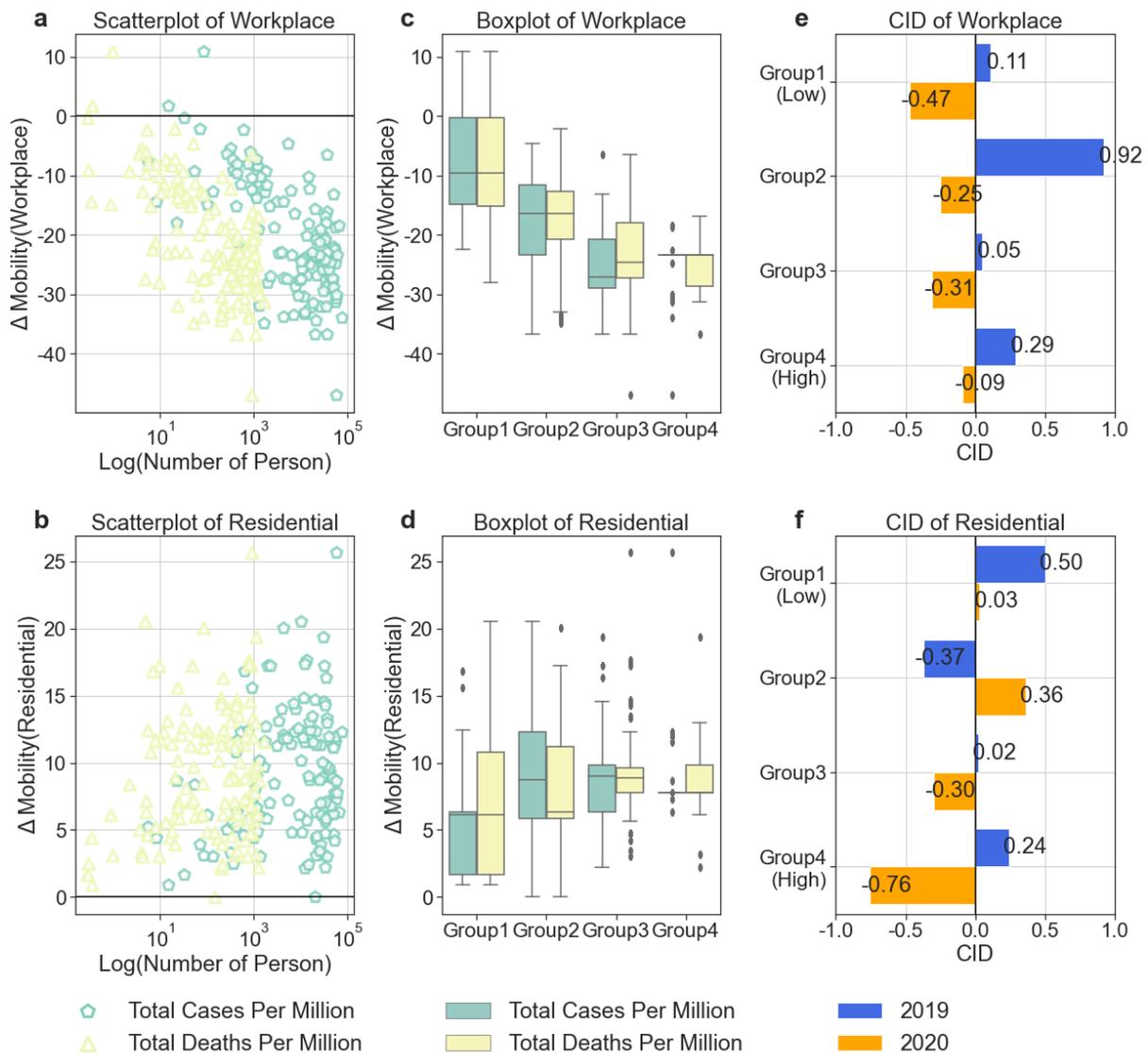

**Figure 4. Distribution between COVID-19 severity and mobility and CID (%) for female first authors by mobility.** The mobility score indicates how visits and lengths of stay at certain locations changes over time compared to a pre-pandemic period (see Methods; [18]). Panels **a** and **b** show correlation between severity of COVID-19 and mobility by country. The x-axis is the logarithm of total cases per million (green) and total deaths per million (yellow). Panels **c** and **d** show the change in mobility by the severity groups as of Fig. 3. Panels **e** and **f** display CIDs for female first authors by their change in mobility. We grouped authors into four levels to estimate how lockdown affects research activity during the pandemic (for CID between 2018 and 2019 compensating for yearly fluctuation, see Supplementary Fig. 6).

This hypothesis suggests that the impact of COVID-19 severity may have indirectly affected female research productivity by changes in national-level distancing or lockdown policy. We can test this mediated effect by examining the extent of change in female productivity according to changes in national mobility. As we mentioned in the previous sections, overall research productivity increased, although the proportion of female leading authors did not keep up with the increased opportunity post-pandemic in 2020. We hypothesise that the correlation between COVID-19 severity and research productivity of female first authors is largely determined by the correlation between severity and change in mobility. In other words, the research productivity of female first authors decreases when they travel less to their workplace and spend more time in their homes.

We observe that workplace mobility decreases when total deaths/cases per million increases (Figs. 4a and c). However, residential mobility increases when total cases/deaths per million increases (Figs. 4b and d). It appears that workplaces closed and time spent at home increased in countries that implemented a strict policy as COVID-19 became more severe. As we hypothesised, research productivity decreased less with more time spent at work (Fig. 4e). Compared to other groups, CID in 2020 decreased the most in group 1 (=−0.47). The CID for group 4, who spend more time in the workplace, decreased least in 2020 (=−0.09). By contrast, as residential mobility increases, research productivity decreases, except for group 2 (Fig. 4f). CID for group 1 in 2020 (=0.03) is bigger than group 4 (=-0.76). In the case of groups 1, 3, and 4, CID in 2020 decreases when residential mobility is higher.

*Relative Significance of Factors for Change in Productivity*

So far, we have explored factors related to the decline of female first authorship. Our finding of multiple factors prompts two key questions: is the decline sincerely related to the pandemic and is any single factor more important than the others. The first question can be answered by the high statistical significance indicated by the p-values being almost zero for all negative CIDs, which confirms our observations in the figures (see Supplementary Table 1 and Methods). However, the p-value does not indicate the degree of influence for each factor. To address the remaining second question, we performed a binary classification for the presence of a female author for each author role by constructing an ensemble of the tree models using XGBoost (see Methods). We compared feature importance (between 2019 and 2020) by the role of female authors (Fig. 5). We used four covid-unrelated features for 2019 and added two covid-related features for 2020. The two covid-related features were workplace and residential mobility—shown to have a significant impact on research productivity for female authors (Figs. 4e and f).

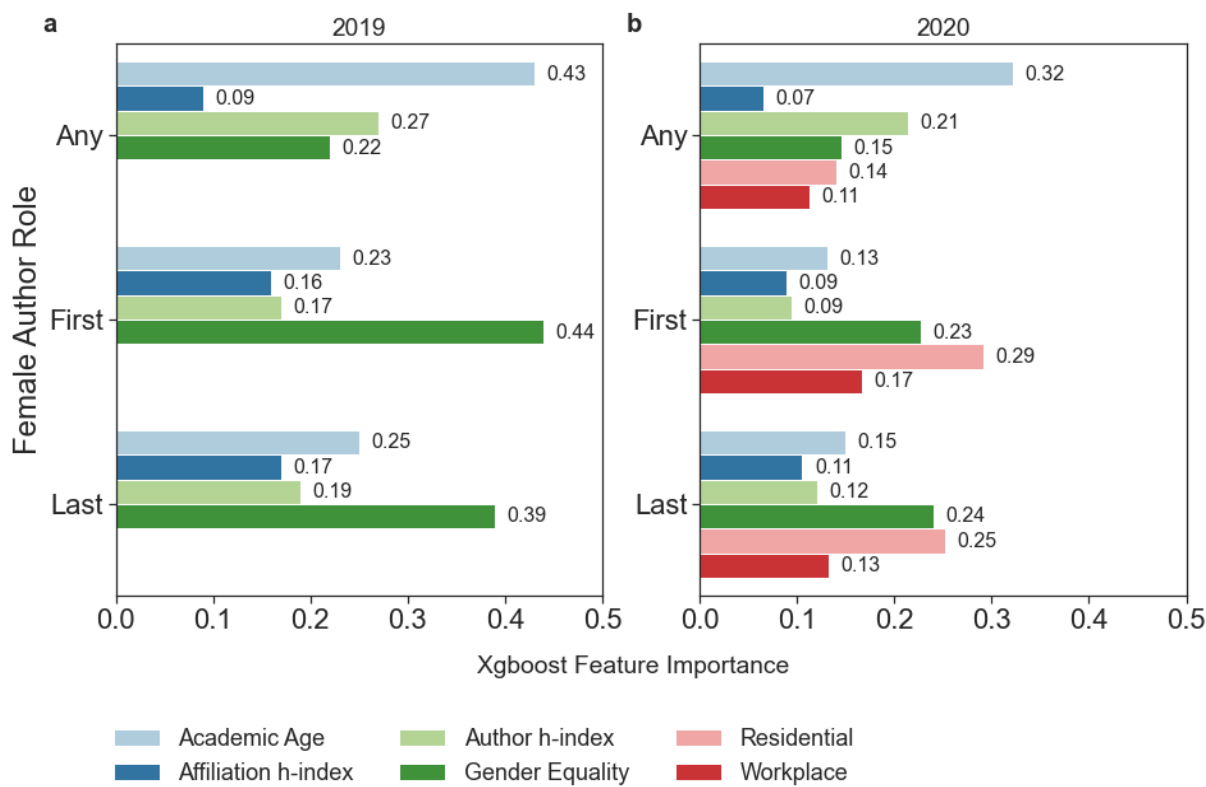

**Figure 5. Feature importance by female author's characteristics measured with XGBoost.** We present the f-score denoting the relative feature importance for the existence of female authors by their role at **a,** 2019 and **b,** 2020, respectively. This calculates the feature importance through the reduced average training loss when using the feature by the gain option of XGBoost (see Methods). We normalise the f-score by dividing the score by their total sum for a fair comparison. We yielded acceptable accuracy for all six models. Specifically, the model accuracies are 0.694 (any female author), 0.655 (female first author), and 0.739 (female last author) for 2019, whereas those in 2020 are 0.708, 0.649, and 0.733, respectively. Note that only four features are used for 2019 because it was before the onset of the pandemic, whereas we additionally used mobility features for 2020. To check the robustness of our model, we also tested the models that include all features and four author characteristics along with residential mobility, with similar results (Supplementary Figs. 7a and b). Compared with the other features, gender equality and COVID-related mobility at the national level are more important for predicting whether the leading authors (that is, first and last authors) are female or not.

Both pre- and post-pandemic, having a female author at any role in a paper is mainly influenced by the average academic age of authors. Over time, more female researchers have entered academia [42]; thus, there is a greater possibility of including female authors when the team members are young. However, the most important feature in 2019 was gender equality among leading authors (Fig. 5a), although it became the second important feature for leading authors in 2020. The most important feature in 2020 switched to residential mobility (Fig. 5b). Therefore, our classification results support the importance of covid-related constraints at the national level that prevent female researchers from leading research during the pandemic. Consequently, it would appear that these constraints are more important than a researcher's career stage and status at the individual and organisational levels.

## Discussion

Although gender equality has gradually improved, a divide still persists in which gender roles impose housework and childcare predominantly to females, which essentially disrupts research productivity of female researchers [9,13,14]. Previous studies have reported the impact of COVID-19 on female research productivity, although these were usually based on partial observations of specific journals and disciplines, leaving many questions unanswered. In this study, we examined the change of female researchers' productivity using a massive collection of bibliographic metadata that spans all academic fields (Microsoft Academic Graph). We did not observe a notable decline in just the proportion of female authors but found a sharp decline of female leading authors, whether first or last authors, at the global scale during the pandemic. The risk arising from the pandemic may have been so subtle that female researchers have not necessarily been excluded from publications but have been marginalised in the publication process.

The global decline of leading female researchers calls for attention from the academic community, particularly considering the pre-pandemic year when the CIDs of female leading authors had been positive in many fields. In short, the outbreak of COVID-19 at best nullified and at worst reversed the recent growth in overall influence of female authors [42,46]. Our analysis on how the decline varied with the academic and social status revealed the more severe damage to the weeks and minors in academia than established researchers.

First, we observed that mid-career female scholars experienced the greatest decrease in productivity as the first author arguably because they are likely to have minor offspring. We also found that the decline in productivity is steeper when the authors are in a relatively unstable academic position (measured by their h-indexes regarding themselves and their institutional affiliations).

At the same time, we found strikingly sharp effects of gender inequality and pandemic-related risks at the national level. First, the greater gender inequality in professional occupations, the greater decline in female first authorship during the pandemic. Second, female first authorship was discouraged by higher COVID-19 mortality and restricted mobility by social distancing. Taken together, these findings suggest that pandemic situations unfavourable to female researchers may be amplified by their disadvantages in the national job market and by quarantine policies in their country of domicile.

Our machine-learning-based analysis also suggests that those national-level factors play more important roles than individual and organisational factors, such as academic age and h-indices for determining female researchers' leading authorships. This finding leads to an interesting policy implication for the global academic community. According to our findings, policies to prevent female researchers from being marginalised during the pandemic may target those in their mid-career with few citations yet, and who work in less influential universities and research organisations. At the same time, our findings suggest that more effective than those individually targeted supports could be global efforts to improve countries with severe gender inequality in the professional labour market and those subject to intense social distancing due to high epidemic mortality. Targeting of these countries does not directly support female researchers in need but may eventually prove beneficial.

The implication of our study is not limited to a report of the current state of academia during the pandemic but can be expanded to the future direction of academia as it recovers in the post-pandemic age. For instance, our study shows that the negative effect of the pandemic on female research leadership is not limited to a few specific journals or disciplines; furthermore, the decrease in female first authorship was present for most countries, except those that have been able to control epidemic mortality. Note that our analysis is based on a collection of socio-academic data to perform pan-academic–scale analysis, rather than using a survey methodology. Such data cannot completely capture each individual's environment, such as the division of housework in a household and the status of offspring. Despite this lack of complete information, we observed significant collective tendencies supporting our conclusions; that is, weaker and minor female researchers were more adversely affected during the pandemic.

Despite an increase in research participation of female authors over time, there are still disadvantages in taking leading roles, as we demonstrated with first authorship during the pandemic. This is consistent with the fact that female authors are more likely to be devalued from authorship as in previous studies [26–28]. We also demonstrated that the authorship of female researchers is affected by the larger social status of labour market equality and work-from-home arrangements. Diversity is key to academic progress [47]. Our study asks academia to pay attention to biased social structures that can unintendedly hinder diversity and suggests that vital actions are taken to protect vulnerable researchers against bias, specifically female researchers who lose opportunities for academic leadership.

## Methods

### *Data Description*

We used the February 15, 2021 dump of the Microsoft Academic Graph (MAG) dataset in this study [48]. This dataset includes bibliographic information of scientific items, such as journal papers, patents, and conference proceedings, in tab-separated values (TSV) format, with their metadata. Additional socio-economic data were collected from various sources, including the gender equality index and total infected cases with mortality [40,41]. Specifically, we used the LinkedIn gender gap indices (GGI), which calculate the difference between genders in their occupations [40]. A higher value indicates a more balanced country regarding gendered occupations, with an average of 0.73 among 170 countries. The total of infected cases and mortality of COVID-19 were retrieved from the COVID-19 Map provided by the Johns Hopkins Coronavirus Resource Center [41]. We used cumulative statistics to December 31, 2020. The average of total cases per million was 16051.62 across 174 countries, and total deaths per million averaged 311.94 across 163 countries. We also used Covid-19 Community Mobility Reports from Google to retrieve the differences in mobility during the pandemic [18]. This data has six place categories from which we selected two relevant places (Workplaces and Residential mobility). Each mobility was calculated for an average between the first date of report to December 31, 2020. A lower score of mobility means less movement at their place compared with the pre-pandemic period. The average workplace mobility was −20.47 (134 countries) and residential mobility was 9.14 (131 countries).

To construct the test dataset, we selected papers published between 2016 and 2020 (see Table 1 for the statistical details). We first filtered all items published during a period of 5 years (2016–2020) from the *Date* element. We then filtered only journal papers that denote the date of an offline publication. We additionally analysed online preprints, where no peer-review had been completed (see Supplementary Figs. 8–12). Because they are not officially published, we used the *OnlineDate* element for the document release date in preprint services. There may be duplicates that have two or more distinct records of a published paper and preprint; however, this does not influence the results because we separated two sets. The conventional practices regarding the preprint are different among disciplines. For instance, one may release the preprint at the time of the submission, meanwhile, others may upload after journal publication. We filter out papers with 50 or more authors to focus on the general practice of science and technology because the ecosystem of the large sciences is different from the typical sciences. After filtering, the average number of co-authors was reduced from 4.8 to 4.5.

|  | N (%) | Mean | SD | Min | Max |
|---|---|---|---|---|---|
| *Papers* | | | | | |
| Number of Journals | 36,005(100) | | | | |
| Number of Issues | 862,193(100) | | | | |
| Number of papers | 15,280,382(100) | | | | |
| Number of coauthors | | 4.55 | 3.61 | 1 | 50 |
| Number of gender-identified papers | 15,092,866(100) | | | | |
| Number of papers with any female author | 9,772,377(64.75) | | | | |
| Number of papers with a female first author | 5,070,922(34.59) | | | | |
| *Individual characteristics by the first author* | | | | | |
| Academic Age | 11,794,302(78.14) | 9.75 | 9.88 | 0 | 60 |
| Author h-index | 11,794,046(78.14) | 5.86 | 7.33 | 0 | 146 |
| Affiliation h-index | 11,794,031(78.14) | 167.40 | 118.14 | 0 | 674 |
| *National characteristics by country* | | | | | |
| Gender Equality | 170(100) | 0.73 | 0.32 | 0.13 | 2.12 |
| Total Cases Per Million | 174(100) | 16051.62 | 18720.89 | 5.64 | 76818.85 |
| Total Deaths Per million | 163(100) | 311.94 | 384.66 | 0.17 | 1684.96 |
| Workplace Mobility | 134(100) | -20.47 | 9.43 | -46.98 | 10.92 |
| Residential Mobility | 131(100) | 9.14 | 4.67 | 0.03 | 25.64 |

**Table 1. Descriptive statistics of the bibliographic data set.** We analysed the metadata in a paper unit, and thus we present the paper-level statistics. The researcher's characteristics were categorised into individual characteristics and country-level characteristics. We present the number of papers for individual-level characteristics. We also display the number of countries with country-level statistics because they were gathered incompletely from the original sources.

### *Identification of Authors and Affiliations for a Paper*

To analyse the scientific activities of authors and affiliations, it is necessary to identify the identities of authors and affiliations. For identification, we used the *Paper Author Affiliations* table in Microsoft Academic Graph. This table contains columns named AuthorId and AffiliationId, which are unique identifiers of the authors and affiliations, respectively, along with the unique identifier of the paper (*PaperId*). We used these identifiers for other analyses. The additional information of the paper was also extracted from the other tables in Microsoft Academic Graph (such as the country information from the *Iso3166Code* column in the *Affiliations* table, which documents two-letter country codes defined in ISO3166 Standards). We disregarded the papers where the affiliations were not assigned to the authors for the analysis using the country information. In addition, we defined the academic age as the elapsed time between the author's first journal paper and the present. Author sequences were retrieved from the *AuthorSequenceNumber* column in the *Paper Author Affiliations* table to identify the first and last authors of a paper. It could be argued that the importance of the first author may vary by discipline. For example, it has been noted that authorship is presented in alphabetical order for mathematics. However, there is no significant difference in the fraction of the alphabetically ordered papers across the disciplines (Supplementary Figs. 13 and 14); thus, we consider the first authors as the leading authors regardless of the FOS.

### *Estimation of the h-index for Authors and Affiliations*

To capture the prestige of authors and affiliations, we used the h-index: a well-known citation-based metric [36] defined as the largest number of h that the given author (affiliations) has published h papers cited at least h times each. Thus, it needs to estimate the number of citations for all papers. For this purpose, we extracted all forward citations from the reference list in the *Paper References* table. We then calculated the citation count for each paper to estimate the h-index by assigning the papers to authors and affiliations. For the h-index, we used MAG regardless of the publish year.

### *Gender Disambiguation*

We used Genderize (https://genderize.io/) to assign the gender of authors. Genderize is a paid web application using statistics collected from the user profiles of social networks and census data. We then inferred gender from the authors' first name, which is the first token of the full name split by space. Because it is based on statistics, the performance of name-based identification cannot be perfect. Low accuracy has been reported for Asian countries, such as China, Japan, and South Korea [49]. Because there are 21,804,885 unique authors in the dataset, it is almost impossible to identify their gender manually. Genderize is also known to be more accurate than other identifiers [49], which was why we selected it for our analysis.

### *Change in Difference Analysis*

For each group, we used CID analysis to verify the increase or decrease in female authors' productivity compared to males for each year (see Table 2). Considering academic lifespan, we grouped academic ages into ten-year groups. Because the h-index is heavy-tailed, where 50% of authors are 1 or less, we set intervals 2 to the power of n. Other variables were grouped into four sets based on their quartiles. The CIDs were estimated in the following way. Let the total number of papers in year $X$ be $A(X)$ and the number of papers by female authors be $B(X)$. Then we can estimate the part of female authors by $F(X) = B(X)/A(X)$, which is the proportion of papers with a female author. Therefore, the CID is defined as follows:

$$CID(X) = [F(X) - F(X-1)] - [F(X-1) - F(X-2)]$$

|  | Group | | | |
|---|---|---|---|---|
|  | Group1 | Group2 | Group3 | Group4 |
| *Individual Characteristics* | | | | |
| Academic Age | [0, 10) | [10, 20) | [20, 30) | [30, ∞) |
| Author h-index | [0, 2) | [2, 4) | [4, 8) | [8, ∞) |
| Affiliation h-index | [0, 61) | [61, 117) | [117, 202) | [202, ∞) |
| *National Characteristics* | | | | |
| Gender Equality | [0, 0.53) | [0.53, 0.85) | [0.85, 0.93) | [0.93, ∞) |
| Total Cases Per Million | [0, 1114.71) | [1114.71, 21012.62) | [21012.62, 52260.65) | [52260.65, ∞) |
| Total Deaths Per million | [0, 14.55) | [14.55, 417.62) | [417.62, 1063.08) | [1063.08, ∞) |
| Workplace Mobility | [0, -24.61) | [-24.61, -23.31) | [-23.31, -16.43) | [-16.43, ∞) |
| Residential Mobility | [0, 6.99) | [6.99, 7.76) | [7.76, 9.63) | [9.63, ∞) |

**Table 2. Group ranges for author characteristics.** We divided the papers into four groups by their characteristic measures. A group range denoted as [a, b) indicates the group contains authors whose characteristic measure $x$ is $b > x \geq a$. For the analysis of the first authors, we used the characteristics of the first authors. Similarly, we used the last authors' statistics when we accounted for the last authors. However, the average was used for the other authors and all authors.

*Two Proportion Z-Test and Combining P-values*

The existence of a female author in a paper can be considered a single random variable of the Bernoulli process; thus, the ratio essentially follows the binary distribution. For a large number of samples, the binary distribution can be estimated as the Gaussian distribution. Thus, to yield the p-values for the CIDs, we performed a two-proportion z-test under the null hypothesis (H0) when the CIDs are less than 0 in 2020 as follows:

$$H0: CID_{2020} = Z_{2020} - Z_{2019} > 0$$

In this case, the test statistic $Z_t$ can perform a one-tailed z-test and can be written as follows by the summation property of the Gaussian variables:

$$\hat{p}_0 = = \frac{\hat{p}_t m + \hat{p}_{t-1} n}{m+n}, \ Z_t = \frac{(\hat{p}_t - \hat{p}_{t-1})}{SE_t} = \frac{(\hat{p}_t - \hat{p}_{t-1})}{\sqrt{\hat{p}_0(1-\hat{p}_0)(\frac{1}{m}+\frac{1}{n})}},$$

$$CID_{2020} = \frac{(Z_{2020} - Z_{2019})}{\sqrt{SE_{2020}^2 + SE_{2019}^2}}.$$

Here, $\hat{p}_0$ is overall sample proportion and $m$ and $n$ are sample sizes at time $t$ and $t-1$. Term $\hat{p}_t$ ($\hat{p}_{t-1}$) is the sample proportion at $t(t-1)$; thus, $SE_t$ denotes the standard error at time $t$.

Because there are four groups for each characteristic, we need to yield a single statistic to validate our results. Fisher's method combines multiple p-values to obtain a single p-value to confirm all groups are statistically significant when the groups are statistically independent [50]. We performed a chi-square using Fisher's methods as follows:

$$X_{2k}^2 \sim -2 \sum_{i=1}^{k} \ln(p_i),$$

where $p_i$ is the *p*-value for each group $i$.

*Estimation of Feature Significance Using XGboost*

*XGboost* is a scalable tree boosting system widely used in machine learning tasks [51]. We tested the feature importance of variables to predict the existence of female authors by published year and author role. Because it was pre-pandemic, we excluded the total infected rate and mortality rate for 2019. Meanwhile, we used six features for 2020. We used the characteristics of the first authors to model the existence of the first authors, whereas we used the last authors' statistics when we accounted for the last authors. However, the average was used for the other authors and all authors.

We calculated feature importance through the reduced average training loss when using the feature by using the gain option of XGBoost. We normalised the f-score by dividing the score by their total sum for a fair comparison. We used the standard train/test splits with a ratio of 9:1 and then trained the model with the following parameters: n_estimators = 1000, learning _rate = 0.1, max_depth = 7.

## Data availability

Bibliographic metadata of academic papers were retrieved from the Microsoft Academic Graph (https://academic.microsoft.com/; https://aka.ms/msracad) on Azure Storage (https://azure.microsoft.com/) licensed under ODC-BY. The gender disambiguation data set is owned by Demografix ApS and can be accessed from their website by means of a subscription (https://genderize.io/). Additional socio-economic data sets are also available from the following sources: epjds-professional-gender-gaps (https://github.com/fverkroost/epjds-professional-gender-gaps) for gender equality; CSSE (https://github.com/CSSEGISandData/COVID-19) for COVID-19 statistics; COVID-19 Community Mobility Reports (https://www.google.com/covid19/mobility) for mobility.

## Code availability

The codes used to analyse and create the figures are available from GitHub (https://github.com/EunrangKwonRepository/MAG).

**Acknowledgements**

This research was supported by the BK21 FOUR (Fostering Outstanding Universities for Research) Program funded by the Ministry of Education of the Republic of Korea and the National Research Foundation of Korea (E.K., J.K.). The National Research Foundation (NRF) of Korea Grant funded by the Korean Government also supported this work through Grant No. NRF-2020R1A2C1100489 (J.Y.). The Korea Institute of Science and Technology Information (KISTI) also supported this work by providing KREONET, the high-speed internet connection. The funders had no role in the study design, data collection and analysis, decision to publish, or preparation of the manuscript.

**Author contributions** All three authors designed the research and wrote the paper. Eunrang Kwon collected and analysed the data.

**Competing interests** The authors declare that they have no competing interests.

**Supplementary information**

Supplementary Figs. 1–14 and Table 1.